\documentclass[prl,groupedaddress,twocolumn]{revtex4}
\usepackage{amsmath}
\usepackage{graphicx}% Include figure files
\usepackage{dcolumn}% Align table columns on decimal point
\usepackage{bm}% bold math
\usepackage{alltt}% 
\usepackage{asymptote}
\usepackage{color,soul}
\definecolor{orange}{rgb}{0.1,0.99,0.95}
\sethlcolor{orange}

\newcommand{\mX}{\mathcal{X}}

\begin{document}

%\title{Support for the histone code hypothesis in an explicit relation 
%between the binding free energy and binding site geometry in a multivalent system}
\title{The entropic lock and key of the histone code}
%Statistical mechanics of the histone code
\author{Bradley M. Dickson}
\email{bdickson@ad.unc.edu}
\affiliation{Center for Integrative Chemical Biology and Drug Discovery, Division of Biology and Medicinal Chemistry, University of North Carolina Eshelman School of Pharmacy, University of North Carolina at Chapel Hill, Chapel Hill, North Carolina, USA.}
\author{Dmitri B. Kireev}
\email{dmitri.kireev@unc.edu}
\affiliation{Center for Integrative Chemical Biology and Drug Discovery, Division of Biology and Medicinal Chemistry, University of North Carolina Eshelman School of Pharmacy, University of North Carolina at Chapel Hill, Chapel Hill, North Carolina, USA.}

%\begin{document} 
\date{\today}

\begin{abstract}
The intricate pattern of chemical modifications on DNA and histones, the ``histone code'', is considered to be a key gene regulation factor. Multivalency is seen by many as an essential instrument to transmit the ``encoded'' information to the transcription machinery via multi-domain effector proteins and chromatin-associated complexes. However, as examples of multivalent histone engagement accumulate, an apparent contradiction is emerging. The isolated effector domains are notably weak binders, thus it is often asserted that the entropic cost of orienting multiple domains can be ``prepaid'' by a rigid tether. Meanwhile, evidence suggests that the tethers are largely disordered and offer little rigidity. Here we consider a mechanism to ``prepay'' the entropic costs of orienting the domains for binding, not through rigidity of the tether but through the careful spacing of the modifications on chromatin. An all-atom molecular dynamics study of the most fully characterized 
multivalent chromatin effector conforms to the conditions for an optimal free-energy payout, as predicted by the model discussed here. 

\end{abstract}

\maketitle

In eukaryotic cells, DNA is wrapped around histone octamers to form nucleosomes and is then packaged into chromatin, a higher order structure. The tails of the histones extend out beyond the DNA and display numerous post-translational modifications (PTMs)\cite{ptms}. Along with histones, nearly 30 additional proteins per nucleosome are associated with chromatin and there are tens of millions of nucleosomes per nucleus\cite{bigpic}. Some of these additional proteins - "effectors" - can bind to the histone PTMs and consequently regulate gene expression by impacting the shape and function of chromatin. The hypothesis that patterns of PTMs transmit biologically relevant signals through recruitment of specific binding partners is referred to as the histone code\cite{strahl,sr4}.

Multivalency was invoked within the histone code as a way to \textit{(i)} overcome the promiscuous binding of different PTMs exhibited by isolated effector domains, \textit{(ii)} to enhance binding relative to that of individual domains, and \textit{(iii)} to maintain a mechanism for rapid response to a change in cellular signaling or environmental stress.\cite{ruthallis,kme4review} Since then, several studies demonstrated the possibility of multivalent chromatin-effector interactions in designed \textit{in vitro} systems\cite{sr9,sr10,sr12}, but it was not until very recently that such interactions were revealed to exist in vivo and be responsible for cellular effects\cite{sr11,scott2012,sr2013}. 
In these instances, multiple structured domains connected by flexible disordered tethers allow the effector to enjoy multivalent engagement with chromatin.\cite{kme4review,ruthallis,pocket,canvas} One example of a multi-domain effector is the E3 ubiquitin ligase UHRF1. The linked tandem tudor domain (TTD) and plant homeodomain (PHD) of UHRF1 are known to engage the H3 histone tail in a multivalent interaction.\cite{3ask,3ask2} The TTD-PHD cassette of UHRF1 is the most well-characterized example of multivalency in the histone code, complete with structural coordinates of the bound state\cite{3ask} and an \textit{in vivo} demonstration that the multivalent bound state is the biologically significant state\cite{sr2013}. 

While some initial effort was made to provide a qualitative depiction of the histone code thermodynamics, multiple questions still remain unanswered. For instance, multivalency was invoked in part to compensate for the weak binding of isolated effector domains. If multiple domains are connected by a flexible tether then some entropic penalty must be levied against the free energy of multivalent binding and this could be significant. To save the situation, one can suppose the tethers are rigid.\cite{ruthallis, kme4review} The problem is that the inter-domain tethers found in multidomain chromatin effector proteins are almost unanimously identified as disordered by a consensus of several reliable techniques\cite{iupred}.

In what follows we suggest an alternative to this conundrum of entropy compensation that does not rely on tether rigidity. Instead, we hypothesize that the spacings of PTMs on chromatin are ``paired'' to an effector so as to minimize the entropic penalty associated with orienting the effector protein domains --- resulting in an entropic lock and key. Multivalent interactions have been studied extensively outside of the chromatin context\cite{jencks,kitov,krish,knapp,saiz} but the problem at hand is unique in that we explore the possibility that selectivity in the histone code is driven, in part, by the minimization of entropic losses upon chromatin binding via PTM spacing. 
In this view, effector proteins act as a ``feeler gage'' that selectively identify sets of PTMs spaced at particular intervals, a view which highlights the importance of the histone sequences which carry the PTMs as those sequences ultimately determine this spacing. 

This alternative means of paying the entropic costs is discussed below through a simple model of the TTD-PHD cassette of UHRF1, wherein the TTD and PHD are non-interacting points connected through a third point representing the inter-domain tether. Given the weak binding enthalpies of the individual effector domains, we suggest that there are two possibilities for designing the chromatin-effector pair. One possibility is that the tether is designed to optimize the free energy of binding for a given chromatin site. The other possibility is that the binding site is such that the free energy of binding is optimal given an inter-domain tether --- we refer to this situation as the entropic lock and key because, as we show below, the entropic penalty of orienting the two binding domains has been minimized. These two possibilities place different physical constraints on the geometry of the effector-chromatin pair and may imply different evolutionary pathways or functional contexts. For example, {\it cis-} and {\it trans-}histone binding may differentially exhibit these two possibilities.
%
%When comparing the simple model to simulation data of TTD-PHD and its chromatin target, we do not find evedence that the inter-domain tether was optimized for the chromatin target. Rather, we find that the chromatin target presents the optimal binding site for TTD-PHD --- 

Our model consists of two particles 
in a box with edge lengths $l$ and volume $l^3$ and we limit ourselves to a special case of multivalency, a divalent binding event. 
The particles $p_1$ and $p_2$ represent two effector protein domains. 
We impose two binding 
sites in the box, located at points $S_1$ and $S_2$, and we ascribe to 
each a radius of interaction $R_1$ and $R_2$ respectively. The binding sites are separated by a distance $D$. 
In this model, a binding event occurs 
if the effector $p_i$ is within $R_i$ of $S_i$. The $i$-particle cannot bind the $j$-particle binding site. 

In this 
system, the probability of finding $p_1$ bound to $S_1$ and $p_2$ bound to 
$S_2$ is straight forward, %when neither particle is coupled to the tether,
\[ P_0 = 
\frac{\int_{\mX}\int_{\mX}1_{[|p_1-S_1|\leq R_1]}1_{[|p_2-S_2|\leq R_2]}dp_1 dp_2}
{\int_{\mX}\int_{\mX}dp_1 dp_2}, \] 
where $\mX$ is the configuration space contained in the volume $l^3$, $1_{[\cdot]}$ are indicator functions which are equal to unity when the argument is satisfied and zero otherwise. We use $|p_i-S_i|$ to indicate the distance between the points $p_i$ and $S_i$. We also use $\int_{\mX}dy$ to indicate integration over the three dimensional Cartesian space of particle $y$. 

$P_0$ evaluates to 
\begin{equation}\label{ideal}
P_0 = \frac{4}{3}\pi\left(\frac{R_1}{l}\right)^3 \frac{4}{3}\pi\left(\frac{R_2}{l}\right)^3. 
\end{equation} 
The probability of a ``dual-binding'' event is simply the product of the probabilities that 
$p_1$ is bound to $S_1$ and that $p_2$ is bound to $S_2$. 
We take this as our reference for measuring the impacts of tethering the two 
particles together so that binding $S_1$ and $S_2$ becomes a divalent affair. 
Note that the free energy of binding is $\Delta G_0= -k_BT\ln(P_0)$ where $T$ 
is temperature and $k_B$ is Boltzmann's constant. 

Suppose we introduce a third particle, $p_{12}$, that serves as a ``tether'' and results in a three particle coarse graining of a dual-domain effector protein. We couple $p_2$ and $p_1$ to the tether with a potential energy function that is zero if $|p_i-p_{12}|\leq L/2$ and infinity if $|p_i-p_{12}|> L/2$ for $i=1,2$. The two domains are thus always together, within a sphere of radius $L/2$ and we have an effective tether length of $L$. We refer to this sphere as the \textit{restraint sphere}. 

Formally, the configuration integral for this coupled system is
\begin{widetext}
\begin{equation} \label{config}%\begin{split}
P=
\frac{\int_{\mX}\int_{\mX}\int_{\mX}1_{[|p_1-S_1|\leq R_1]}1_{[|p_2-S_2|\leq R_2]}1_{[|p_1-p_{12}|\leq L/2]} 1_{[|p_2-p_{12}|\leq L/2]}dp_1 dp_2 dp_{12}}
{\int_{\mX}\int_{\mX}\int_{\mX}1_{[|p_1-p_{12}|\leq L/2]}1_{[|p_2-p_{12}|\leq L/2]}dp_1 dp_2 dp_{12}}. 
%\end{split}
\end{equation} 
\end{widetext} 

We make the following arguments under the approximations: \textit{i)} collisions between the particles and the walls of the container may be ignored so long as the center of the restraint sphere, $p_{12}$, is inside the container, \textit{i.e.},$L << l$, 
and \textit{ii)} we consider only whole volumes of the binding sites even when the restraint sphere technically only covers a fraction of the volume. 

Under these approximations, we can construct the probability of observing the dual bound state $P$ (Equation \eqref{config}) as the product of three probabilities: The first factor is the probability that the restraint sphere 
containing $p_1$ and $p_2$ also contains $S_1$ and $S_2$. 
The second factor is the probability that 
$p_1$ is in $S_1$ given that the restraint sphere contains the particles and the binding sites. The third 
factor 
is just like the second factor, except it is evaluated for $p_2$ in $S_2$. 
These last two factors are expressed as the ratio of the volume of a given binding site to the volume of the 
restraint sphere.

The first factor of the probability can be deduced as follows. When $p_1$ is at $S_1$, the possible positions of $p_{12}$ trace a sphere of radius $L/2$. When $p_2$ is at $S_2$, $p_{12}$ traces out another sphere of radius $L/2$. If $p_{12}$ is found in the intersection of these two spheres, then divalent binding may be observed. The volume of this intersection can be computed as the sum of two spherical caps. The volume of a spherical cap is $V_c=\pi h^2(3L/2-h)/3$ where $h=(L-D)/2$ is the height of the cap and $L/2$ is the radius of the sphere. By symmetry, the volume of the intersection is $2V_c$. Thus, the ratio of $2V_c$ to the volume of the container is the probability that the restraint sphere contains both binding sites. 
\begin{figure}[h]
\includegraphics[width=3in]{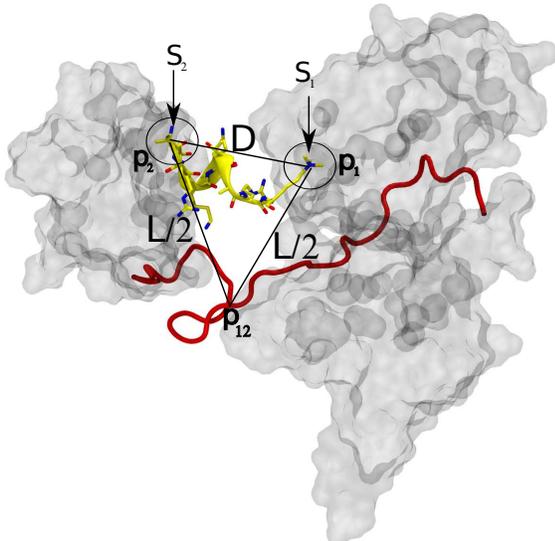}
\vspace{-20pt}
\caption{An illustration of the model with a known multivalent binder UHRF1. The TTD and PHD are shown in clear-spacefill while the inter-domain linker is shown in red. The H3 tail is shown in yellow. 
\label{diag}}
\vspace{-15pt}
\end{figure}

Up to our approximations, the probability of observing the dual-bound state under the influence of this 
tether is  
\begin{equation}\label{coupl}
P = \frac{\frac{32}{3}\pi\left(L+\frac{D}{2}\right)[L-D]^2}{l^3L^6}R_1^3R_2^3. 
\end{equation} An illustration of this model is given in figure \ref{diag}. 
Notice that in this equation the absolute scale is irrelevant and only the ratios of the parameters are important. The approach of Diestler and Knapp\cite{knapp} could be used to arrive at equation \eqref{coupl} but in this case it is easy enough to approximate the configuration integrals directly.

Given the probabilities of the monovalent and divalent binding events, we can now estimate the ``payout'' of divalency
\begin{equation}\label{boost} 
f \equiv \frac{P}{P_0} = \frac{6\left(L+\frac{D}{2}\right)[L-D]^2 l^3}{\pi L^6}. 
\end{equation} 
We can relate the coupled and uncoupled systems with $P= P_0f$ and write the free energy 
of binding as 
\begin{equation}\label{explic}
\begin{split}
\Delta G &= -k_BT \ln(P_0 f)\\
&= \Delta G_0-k_BT\ln\left(\frac{l^3}{L^3}\right)\\
&-k_BT\ln\left(\frac{\left(L+\frac{D}{2}\right)[L-D]^2}{L^3}\right) -\text{const}.
\end{split}
\end{equation}
In this case, by design, the enthalpies of binding events are additive but additivity will not hold in general\cite{jencks} although it is a common approximation in the context of multivalency\cite{kme4review,ruthallis,krish}.

The second term in equation \eqref{explic} 
replaces one entropic contribution of $k_BT\ln(l^3)$ from $\Delta G_0$ with $k_BT\ln\left(L^3\right)$. 
Thus, the full entropy of the two independent 
domains is reduced, effectively leaving one domain to sample the full container while the other domain only samples the restraint sphere. When $l^3$ is large 
this contribution would be as important as the addition of binding enthalpies. 
This term ``prepays'' for the translational freedom of one of the domains that would be lost on binding and is 
responsible for transforming the probability of dual engagement from fleeting to expected. The average mammalian cell nucleus has a radius of about $3 \times 10^4$ \AA \,. If $80 \, \text{to}\, 99\%$ of the nucleus volume is filled, the volume of the empty space is still on the order of $10^9 \,\text{to}\, 10^{10}$ cubic \AA\, suggesting a payout of billions to trillions. 

The third term in equation \eqref{explic} compares the number of ways the restraint sphere can cover the binding sites (the volume of the spherical caps) to the number of ways the two domains can be arranged (the volume of the restraint sphere). This term is the entropic penalty for simultaneously organizing the two domains into their binding sites. As the number of ways to cover the two sites shrinks (as $D$ approaches $L$) this term can completely abolish binding, reflecting a poor fit between the dual-domains and the binding site geometry. 

Equation \eqref{coupl} can be maximized in two ways: {\it i)} the tether length $L$ can be optimized for fixed $D$ or {\it ii)} the binding site distance $D$ can be optimized for fixed $L$. The second case is trivial, when $D=0$ the entropic penalty vanishes. In practice $D$ will be finite due to the excluded volumes of the domains. The first case can be treated by finding the maxima of $(L+\frac{D}{2})(L-D)^2/L^6$ for fixed $D$. The optimal tether length is $L=D/.618$, which causes the domain-domain root-mean-squared separation (RMS) to coincide with $D$. 

These two situations should be easily distinguished by comparing the domain-domain distance distribution with the binding site separation. If the binding site separation is close to the domain-domain RMS, then the tether has been optimized for the binding site. If on the other hand the binding site separation is close to the minimum domain-domain spacing, then the binding site geometry is optimal for binding but the tether has not necessarily been optimized for binding.
%
%The selectivity of a multi-domain effector protein is therefore partially encoded through the tether via a geometric penalty that is minimal for particular PTM spacings. Selectivity is an anticipated feature of multivalent systems in the histone code\cite{kme4review,ruthallis,pocket}, but has not yet been given explicit form. In equation \ref{explic}, we have obtained an explicit shape dependence of the binding free energy that is entropic in origin --- an entropic lock and key. While this particular tether-model may not be representative of all effectors, it is clear that a similar entropic penalty will be a component of the binding interaction for any effector-chromatin pair demonstrating a high degree of conformational plasticity.

We were curious how effectors of the histone code might be paired with their chromatin targets, so we compared the dynamics of the multivalent effector UHRF1 and its target histone H3 with the model captured by equation \eqref{config}. We have combined our previous molecular dynamics (MD) simulations of the TTD-PHD cassette of UHRF1 (nearly $4\mu$s)\cite{sr2013} with a new simulation of the H3 tail and a Monte Carlo (MC) simulation of a 5-particle system obeying equation \eqref{config}. In this example, $p_1$ represents the methyllysine binding pocket of the TTD and $p_2$ represents the N-terminal binding surface of the PHD. The H3 N-terminal nitrogen and the N$^{\zeta}$ if lysine 9 represent the the binding sites $S_1$ and $S_2$ and were used to build the distance distribution $p(D)$. We used the OH atom of Tyr188 and the backbone O atom of Pro355 to build the distance distribution $p(|p_1-p_2|)$ for the TTD-PHD. These choices were based on proximity to the N$^{\zeta}$ and N-terminus, respectively. Details of the TTD-PHD simulation are reported elsewhere\cite{sr2013} and details of the histone simulation are given in Supplemental Materials. 

We computed the free energy along the coordinate $D=|N-N^{\zeta}|$ with an adaptive biasing potential\cite{pre11}.
Error in the free energy along the N-N$^{\zeta}$ distance was estimated as $\int |A(\xi,t_1)-A(\xi,t_2)|d\xi$ for $A(\xi,t)<10$kcal/mol, $\xi=D$, and $t_1=96$ns, $t_2=128$ns. The error in free energy was 1.8 kcal/mol and was confined to values of $D<20$\AA\,. The free energy along $D$ leads to the distribution in $D$ since $p(D)\propto \exp[-\beta A(D)]$.

\begin{figure}[h]
\includegraphics[width=\columnwidth]{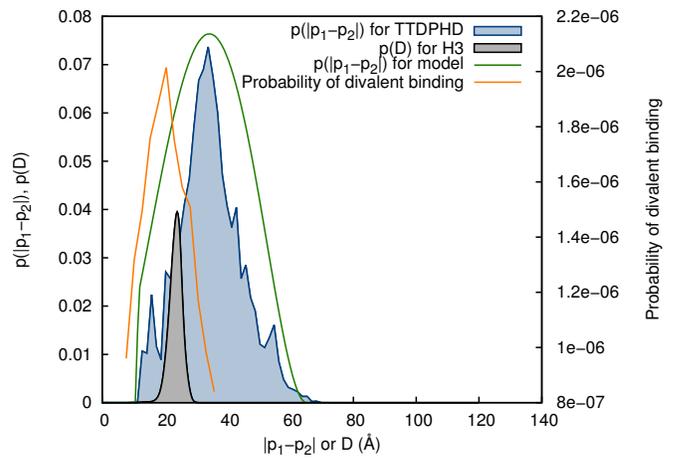}
\vspace{-20pt}
\caption{The distance distribution for the binding sites on histone H3 (black) is shown with the $|p_1-p_2|$ distribution from the TTD-PHD cassette (blue). The probability of divalent binding was evaluated at various $D$ by Monte Carlo (orange). The $|p_1-p_2|$ distribution (green) from the MC simulation is also shown. 
\label{key}}
\vspace{-10pt}
\end{figure}

The distance distribution $p(|p_1-p_2|)$ for the TTD-PHD (shown in blue in Figure \ref{key}) suggests a tether length of $L_{\text{UHRF1}}=65$\AA\, and a particle radius of $6$\AA. We performed Monte Carlo simulations of equation \eqref{config} that includes excluded volume effects where the ratio $6/65$ is used to determine the length scale of the model. We used an excluded volume (or particle) radius of $L \times 6/65$. The MC results shown in figure \ref{key} (orange) were obtained with $R_1=R_2$, $R/L =0.117$, $L/l=0.34$ and various values of $D$. Optimal binding for the model system was observed when $D/L = 0.313$, which suggests an optimal PTM spacing of $D=|N-N^{\zeta}|=0.313 \times 65 = 20.4$\AA\, for the TTD-PHD and H3 system. The distribution of $D$ for H3 is shown in black in figure \ref{config} and peaks near $22$\AA. This is in agreement with the peak in divalent binding probability which occurs at $D=20$\AA. This suggests that UHRF1 is paired with the N-terminus and methylated $N^{\zeta}$ of lysine 9 of H3 because this spacing minimizes the entropic penalty associated with the divalent bound state. Independently of this spacing, binding can be impaired by chemical modification of other histone residues (methylation of Arg2 for example). Thus the entropic lock and key is only one component of the elaborate recognition mechanism that drives the histone code.

It was not possible to infer the parameters $R$ and $l$ from the existing TTD-PHD data. However, $l$ only impacts the magnitude of the probability of observing the dual-bound state and does not shift the optimal $D:L$ ratio. We considered several $R$ and found very little impact on the results for $R/L = 0.06, 0.117, 0.17$. 
%
%In general, either the chromatin targets or the effector protein could be the more rigid of the pair. In the present case the effector is the more flexible of the two but we expect a different situation in \textit{trans}-histone examples of multivalent binding\cite{sr9,sr11}. For the systems studied in references \cite{sr9,sr11} the entropic lock and key hypothesis could be further tested by characterizing the spatial distribution of the chromatin targets and the dual-domains. Models like the one developed here could be built around those mechanical details and used to test the generality of our hypothesis. This kind of work will help guide future experiments beyond an ``all or nothing'' interrogation of multivalency within the histone code and shed light on the basic principles which target effectors to specific chromatin partners.

%\section{Discussion}
In 1894, Emil Fischer suggested an interaction model explaining the enzyme-substrate specified by the complementarity of their respective geometric shapes.\cite{fischer} This type of interaction, generally referred to as ``the lock and key'', was mostly thought to be enthalpy-driven due to multiple contacts resulting from the perfect shape fit. Based on the results presented here, we suggest that an entropic ``lock and key'' might be a fundamental component of design that pairs effector proteins with their chromatin targets. Because the individual modules of multi-domain chromatin effectors are often weak binders to their respective sites and because they are connected by long disordered linkers, multivalent binding of such effectors to chromatin is likely controlled by entropy. %Hence, despite the fact that our model does not account for any binding enthalpies, it still adequately reflects the thermodynamics of the histone code multivalency. 

The physical traits of multivalency resulting from the presented model may have genome-wide implications. For instance, the characteristics of tethers connecting the effector domains, such as length and plasticity, as well as the geometry and dynamics of the target PTMs are an as yet underappreciated component of the histone code. In the case of UHRF1, phosphorylation of the tether in the TTD-PHD cassette is known to weaken histone binding by about 30 fold\cite{3ask} and correlates with enhanced production of TopoII$\alpha$\cite{tethermark}. Additionally, a mutation in the TTD-PHD linker (R295A/R296A) results in an increased tether length and drastically impacts histone binding\cite{3ask} while causing a loss of function\cite{sr2013}. Hence, the function of UHRF1 may be altered by a single modification that changes the properties of the tether between TTD and PHD. Until now, the studies of the histone code effectors were mostly focused on the discovery of new effector domains followed by the elucidation of their structure and function. In light of the presented results, one might suggest that the inter-domain linkers could have been used as an evolutionary speed-dial; By mechanical alteration of a linker, nature could steer an effector away from the areas where its presence might be unwanted, which is less time-consuming and less risky than reengineering the effector domains. 

The present work also provides a means of categorizing effector-chromatin pairs. The TTD-PHD of UHRF1, which binds {\it cis-}histone, appears to bind chromatin opportunistically by selecting chromatin sites with optimal $D$. Will other chromatin effectors demonstrate a similar relationship with their binding partners or will some of them display optimized tethers? These would be two fundamentally different classes of effectors and would represent a significant difference in effector design. It will be interesting to consider CHD4/5\cite{sr10,sr11} or BPTF\cite{sr9}, which are examples of {\it trans-}histone binders. Will the {\it cis-}/{\it trans-}histone preference correlate with the two modes of optimallity outlined here?

\begin{acknowledgements}
The research described here was supported by the US National Institute of General Medical Sciences; US National Institutes of Health (grant RC1GM090732 and R01GM100919); the Carolina Partnership and the University Cancer Research Fund; University of North Carolina at Chapel Hill.
\end{acknowledgements}

\noindent {\bf Supplemental Materials}
%\begin{section}
%Recently, we reported on the dynamics of the TTD-PHD cassette of UHRF1 and in particular how the selectivity of the dual-domain is derived.\cite{sr2013} For that study, we collected nearly $4\mu$-seconds of simulation data during an adaptively biased free energy calculation. To examine the possibility that the entropic lock and key is present in effector protein construction, we have parsed that data for distances between the N-terminal and methylated lysine binding pockets. Because the biasing potential is proportional to the free energy\cite{pre11}, the densely populated conformations in the unbiased phase space distribution are also the most populated conformations in the biased phase space distribution. The distribution computed from our biased trajectory is thus a reasonable indication of the shape of the domain-domain spacing for the TTD-PHD cassette. 

We used the biasing scheme from reference \onlinecite{pre11} with $b=0.8$ and $c=0.1/\Delta t$ ($\Delta t$ is the timestep) to compute the potential of mean force along the distance from the N-terminal nitrogen to N$^{\zeta}$ of lysine 9 in an isolated H3 histone tail (residues 1-10) using a $128$ ns trajectory. Simulation setup is the same as in reference \onlinecite{sr2013} but here we use the GROMOS96ff force field. 

To enhance sampling of backbone dihedrals of the histone we applied a bias potential along the dihedrals. This bias was derived from a converged free energy computation in the $\phi$, $\psi$, and $\omega$ backbone dihedral angles of a tri- alanine peptide. The $(\phi,\psi,\omega)$ computation was carried out with a recently introduced decomposition method\cite{cl11}. In the end, this provides an angle-specific biasing potential that can enhance rotations along the histone backbone. The converged histograms in $\phi, \psi \text{\,and\,} \omega$ from the tri-alanine simulation were used against the histone with $b=0.3$, $c=0.1/\Delta t$. The effects of this bias on the distribution of dihedral angles was removed through on-the-fly-reweighting\cite{ens10,drivenAB} during the N-N$^{\zeta}$ free energy computation. 
%\end{section}

\bibliography{bibs.bib}

\end{document}